\documentclass[twocolumn]{autart}    

\usepackage{graphicx}          
\usepackage{tabularx}
\usepackage{subcaption}

\usepackage{amsmath}            
\numberwithin{equation}{section} 

\begin{document}

\begin{frontmatter}

\title{An identification method for oscillators with response-dependent inertia\thanksref{footnoteinfo}} 

\thanks[footnoteinfo]{This paper was not presented at any IFAC 
meeting. Corresponding author Y.~Harduf. Tel. +1-(781) 655-1754.}

\author[Technion]{Yuval Harduf}\ead{Hardufy2@gmail.com},    
\author[Technion]{Eyal Setter}\ead{Eyalsetter.es@gmail.com},               
\author[Technion]{Izhak Bucher}\ead{Bucher@technion.ac.il}  

\address[Technion]{Technion, Israel Institute of Technology, Faculty of Mechanical Engineering}  

\begin{keyword}                           
Identification methods; Linear/nonlinear models; Variable-mass systems; Inductor saturation; Hilbert transform.                            
\end{keyword}                             

\begin{abstract}                          
This paper is concerned with identifying the instantaneous modal parameters of forced oscillatory systems with response-dependent generalized inertia (mass, inductance, or equivalent) based on their measured dynamics. An identification method is proposed, which is a variation of the "FORCEVIB" method. The method utilizes analytic signal representation and the properties of the Hilbert transform to obtain an analytic relationship between a system's natural frequency and damping coefficient to its response and excitation signals. The proposed method is validated by comparing the identification results to the asymptotic solution of a simple system with response-dependent inertia and is then demonstrated, numerically and experimentally, for other more complicated nonlinear systems.
\end{abstract}
\end{frontmatter}
\section{Introduction}\label{sec:introduction}
\subsection{Variable-inertia systems}\label{subsec:Variable-inertia systems}
Variable-inertia systems are systems whose inertia term is not constant, but rather depends on time or the state of the system \cite{DynMechSysVarMass}.  Such systems can be found in mechanical and electrical oscillator devices. Various systems acquire or dispose of mass as a function of time: a rocket burning fuel, a vessel being emptied or a rotor winding or unwinding material \cite{Cveticanin2022DynamicsMass}. A relevant sub-category of variable-inertia systems, with which this work is concerned, are systems whose inertia depends explicitly on their state; Examples include: a cable being deployed from a reel, where the suspended mass of the cable depends on the laying reel's rotation, the height of a water column in a free-surface piercing pipe \cite{Pesce2014SystemsPosition} or the radial position of a particle moving on a rotating parabola \cite{Nayfeh1995NonlinearOscillations}. Other examples may come from the field of electric circuits – inductors, who take up the role of inertia in the equivalence between mechanical oscillators and RLC circuits, may change their properties based on the current flowing through them \cite{Oliveri2022NonlinearSurvey}. 
Although variable-inertia systems are not very common, they are not a theoretical concept, but real engineering systems, which require the proper tools to be analyzed. Several works were dedicated to the solution of the forward problem of variable-inertia systems, i.e. – obtaining the system’s response from its equations of motion \cite{Marinca2010DeterminationMethod,Venkatesan1997NonlinearSystems,vanHorssen2006OnMass,vanHorssen2010OnMass}. Due to the nonlinear nature of these systems, solving the problem of identifying the system’s equations of motion from its response requires nonlinear identification methods \cite{Kerschen2006PastDynamics,Brunton2016DiscoveringSystems}. However, most of these methods assume a constant-inertia term and to the best of the authors’ knowledge, for forced variable-inertia systems, the problem of identifying the system’s properties from its forced response has not been fully resolved yet. In this paper, we propose a modification to a non-parametric Hilbert-transform-based approach to nonlinear system identification, allowing the identification of forced systems whose inertia term depends on the state of the system.
\subsection{Theoretical background: Hilbert transform and analytic signal representation}\label{subsec:TheoreticalBG}
Many signals, including the response of vibrating systems $y(t)$, can be represented using slowly varying, real-valued functions – envelope, or, amplitude $A(t)$ and phase $\phi(t)$  \cite{Lyons2011UnderstandingLyons.}:
\begin{equation} \label{eq1.1}
y(t)=A(t)\cos(\phi(t))
\end{equation}
The instantaneous frequency of the signal is defined as the derivative of the phase:
\begin{equation}\label{eq1.2}
    \omega(t)=\frac{d\phi(t)}{dt}
\end{equation}
The analytic representation of a signal $Y(t)$ is a complex function of time. The analytic signal can be written using either the signal’s Hilbert transform (HT) or using amplitude and phase:
\begin{equation}\label{eq1.3}
    \begin{split}
        Y(t)&=y(t)+j\Tilde{y}(t)=A(t)\exp(j\phi(t))\\
        A(t)&=\sqrt{y^2(t)+\Tilde{y}^2(t)}\\
        \phi(t)&=\arctan(\Tilde{y}(t)/y(t))
    \end{split}
\end{equation}
Where $j=\sqrt{-1}$  and $\Tilde{y}(t)$ is the HT of $y(t)$, defined as:
\begin{equation}\label{eq1.4}
    \Tilde{y}(t)=H[y(t)]=\frac{1}{\pi}\int_{-\infty}^{\infty} \frac{y(\tau)}{t-\tau},\d\tau
\end{equation}
The HT can be considered as a filter with unit gain that shifts the phase of all positive frequency components of a signal by $-\pi/2$ and the negative components by  $\pi/2$ [14].
\subsection{The FORCEVIB method}\label{subsec:FORCEVIB}
The method proposed in this paper is based on the FORCEVIB method \cite{Feldman1994Non-linearForcevib}. The FORCEVIB method is concerned with variable-stiffness systems of the following form:
\begin{equation}\label{eq1.5}
    \Ddot{y}(t)+h(y,\Dot{y})\Dot{y}(t)+\omega_n^2(y,\Dot{y})y(t)=\frac{1}{m}x(t)
\end{equation}
where $y(t)$ is the system's response, $h(y,\Dot{y}), \omega_n(y,\Dot{y})$ are the system’s symmetric damping coefficient and natural frequency, respectively, $m$ is the system's constant mass and $x(t)$ is the external excitation (input).

The goal of the FORCEVIB method is to estimate the implicit relationships between a system's modal, i.e. steady-state, parameters and its response amplitude $h(A),\omega_n(A)$, from time-series measurements of the system's response and excitation force. 

To do so, the method utilizes the HT and analytic signal representation. There are two benefits to using analytic signal representation for identification purposes. The first is that when the modal parameters vary slowly compared to the system's response, the differential equations for the response of a system hold for the analytic representation of the response as well: According to the Bedrosian theorem \cite{Bedrosian1963ATransforms}, when the spectra of the modal parameters $\omega_n(t),h(t)$ are lower then, and do not overlap with, the spectrum of the response $y(t)$, then $H[h(t)\cdot y(t)]=h(t)\Tilde{y}(t)$. Utilizing the Bedrosian theorem and the linearity of the HT, we can apply the HT to both sides of equation \ref{eq1.5}, multiply it by $j$ and add it to the original equation to obtain a differential equation in the analytic signal:
\begin{equation}\label{eq6}
    \Ddot{Y}(t)+2h(A)\Dot{Y}(t)+\omega_n^2(A)Y(t)=\frac{1}{m}X(t)
\end{equation}
Separating equation \ref{eq6} into its real and imaginary parts and solving for the modal parameters yields explicit relationships between the modal parameters and the system's input and response signals: 
\begin{equation}\label{eq1.6}
    \begin{split}
        \omega _0^2 &=  \frac{x\Dot{\Tilde{y}} - \Tilde{x}\Dot{y}}
                            {m(y\Dot{\Tilde{y}} - \Tilde{y}\Dot{y})} -
                        \frac{\Ddot{\Tilde{y}}\Dot{y} + \Ddot{y}\Dot{\Tilde{y}}}
                            {y\Dot{\Tilde{y}} - \Tilde{y}\Dot{y}}\\ 
        2h &=   \frac{\Tilde{x}y - x\Tilde{y}}
                    {m(y\Dot{\Tilde{y}} - \Tilde{y}\Dot{y})} -
                \frac{\Tilde{y}\Ddot{y} - \Ddot{\Tilde{y}}y}
                    {y\Dot{\Tilde{y}} - \Tilde{y}\Dot{y}}
    \end{split}
\end{equation}
and the value of $m$ can be estimated by solving a line-fitting problem, as described in \cite{Feldman1994Non-linearForcevib}. 

The second benefit of using analytic signal representation is that the HT of a measured signal can be estimated using digital filtering; While such a filter cannot be realized perfectly, a practical filter with unit gain and $\pi/2$ phase delay in the parts of the spectrum contained in a system’s response can be designed and used to estimate the HT of the system’s response \cite{Lyons2011UnderstandingLyons.}. 

Since equation \ref{eq1.5} describes systems with constant inertia, the FORCEVIB method is not suitable for variable-inertia systems. However; with slight modifications, the same ideas can be applied to variable-inertia systems, which is the objective of this paper.
\section{The proposed method}\label{sec:Method}
\renewcommand{\theequation}{\arabic{section}.\arabic{equation}} 
We propose an identification method for forced variable-inertia systems with symmetric parameters whose equations of motion have the following form:
\begin{equation}\label{eq2.1}
    m(y,\Dot{y})\Ddot{y}+c(y,\Dot{y})\Dot{y}+ky=x(t)
\end{equation}
Where $m(y,\Dot{y}), c(y,\Dot{y}), k$  are the system's inertia, damping, and stiffness coefficients. Notice that the stiffness term is constant in this model. While usually it is convenient to normalize the equations of motion by the inertia term, we will normalize the equation by the constant stiffness term to obtain:
\begin{equation}\label{eq2.2}
    T^2(y,\Dot{y})\Ddot{y}+2\chi(y,\Dot{y})\Dot{y}+y=\frac{1}{k}x(t)
\end{equation}
where $T^2=\frac{m(y,\Dot{y})}{k},\chi=\frac{c(y,\Dot{y})}{2k}$.

The goal of the identification would be to identify the implicit relationships between the modal parameters and the response amplitude, i.e. $T(A),\chi(A)$ as well as the constant term $k$ from time-series measurements of the system's response and excitation signals.
Since the method relies solely on the system's input and output time-series measurements, it can also be used for reduced-order modeling of complex systems. By measuring the input and output of any system, the modal parameters of an equivalent variable-inertia system can be identified and used to derive a reduced-order model. 
While $T(A),\chi(A)$ are useful for the identification process, their physical meaning is not immediately apparent. We will use the following relationships to present the identification results in a way that is meaningful and conforms with widespread conventions (see \cite{Feldman2011HilbertVibration}):
\begin{equation}\label{eq2.3}
    \omega_n^2(A)=\frac{1}{T^2(A)},h(A)=\frac{\chi(A)}{T^2(A)}
\end{equation}
Since the inertia variations affect both the natural frequency and damping coefficients, variable inertia systems exhibit a rich dynamic response, as will be shown in the examples brought in section \ref{Case studies} of this paper.
\subsection{Signal filtering and analytic form estimation}\label{subsec:HVD}
As mentioned in the introduction, the proposed method requires estimating the HTs of the excitation and response signals, as well as their derivatives. While the HT can be realized using a digital filter, measurement noise may cause numerical difficulties when differentiating the signals. To handle measurement noise, using the largest component of the Hilbert Vibration Decomposition (HVD), described in \cite{Feldman2006Time-varyingTransform}, is proposed. HVD is used to decompose a signal into non-stationary components in the form of \ref{eq1.1}. By estimating the instantaneous frequency of a signal, averaging the result, and then estimating the corresponding envelope using synchronous demodulation, the largest non-stationary component of the signal can be extracted and used for further calculations. By using the HVD procedure and taking the signal's largest component we ensure: a. the signal's HT is devoid of smaller, additional components, deterministic and random and b. improved robustness to noise when numerically differentiating the signals \cite{Xia2021ModalStructures}. In appendix \ref{App:Noise}, the method's robustness to noise is demonstrated with simulated noise.
\subsection{Identification of varying modal parameters}\label{subsec:IdentModalPara}
Following the ideas presented in \cite{Feldman1994Non-linearForcevib}, equation \ref{eq2.2} is transformed to analytic form in a similar manner to that presented in subsection \ref{subsec:FORCEVIB}: the HT is applied to both sides of the equation. The left-hand-side is multiplied by $j$ and simplified using the Bedrosian theorem and the linearity of HT. The result is added to the original equation to obtain:
\begin{equation}\label{eq2.4}
    T(A)^2\Ddot{Y}+2\chi(A)\Dot{Y}+y=\frac{1}{k}X
\end{equation}
Equation \ref{eq2.4} is complex. Separating the real and imaginary components, we obtain two equations:
\begin{equation}\label{eq2.5}
    \begin{split}
        \text{Real:}        &\quad {T^2}\Ddot{y} + \chi \Dot{y} + y = \frac{1}{k}x \\
        \text{Imaginary:}   &\quad {T^2}\Ddot{\Tilde{y}} + \chi \Dot{\Tilde{y}} + \Tilde{y} = \frac{1}{k}\Tilde{x} \\ 
    \end{split}
\end{equation}
Solving for $T^2,\chi$:
\begin{equation}\label{eq2.6}
    \begin{split}
  {T^2} &= \frac{{x\Dot{\Tilde{y}} - \Tilde{x}\Dot{y}}}{{k\left( {\Ddot{y}\Dot{\Tilde{y}} - \Dot{y}\Ddot{\Tilde{y}}} \right)}} + \frac{{\Tilde{y}\Dot{y} - y\Dot{\Tilde{y}}}}{{\Ddot{y}\Dot{\Tilde{y}} - \Dot{y}\Ddot{\Tilde{y}}}} \quad (a) \\
  \chi  &= \frac{{\Tilde{x}\Ddot{y} - x{\Ddot{\Tilde{y}}}}}{{2k\left( {\Ddot{y}\Dot{\Tilde{y}} - \Dot{y}\Ddot{\Tilde{y}}} \right)}} + \frac{{y\Ddot{\Tilde{y}} - \Tilde{y}\Ddot{y}}}{{2\left( {\Ddot{y}\Dot{\Tilde{y}} - \Dot{y}\Ddot{\Tilde{y}}} \right)}} \quad (b) \\ 
    \end{split}
\end{equation}
Equation \ref{eq2.6} is the solution to the identification problem. The solution contains the input signal $x(t)$ and the response signal $y(t)$, their Hilbert transforms and the derivatives of both the signals and their Hilbert transforms, all of which can be obtained by measuring the input and response signals. The solution also contains the constant stiffness term, whose estimation method is explained in the following subsection.
\subsection{Identification of the constant stiffness term}\label{subsec:ConstTermEst}
Following the ideas in \cite{Feldman1994Non-linearForcevib}, we make use of the slow variation of the modal parameters (see section \ref{subsec:FORCEVIB}) to assume that over a relatively short duration of time the change in the left-hand-side of equation \ref{eq2.6}(a) is negligible and the parameter $T^2$ can be considered as a constant. The equation can be rearranged as: 
\begin{equation}\label{eq2.7}
    g=-\frac{1}{k}s+T^2
\end{equation}
Where $g=\frac{{\Tilde{y}\Dot{y} - y\Dot{\Tilde{y}}}}{{\Ddot{y}\Dot{\Tilde{y}} - \Dot{y}\Ddot{\Tilde{y}}}}, s=\frac{{x\Dot{\Tilde{y}} - \Tilde{x}\Dot{y}}}{{ {\Ddot{y}\Dot{\Tilde{y}} - \Dot{y}\Ddot{\Tilde{y}}}}}$. Equation \ref{eq2.7} describes a straight line with the slope of $-\frac{1}{k}$ and an intercept of $T^2$. To estimate $k$, a simple line fitting procedure can be employed, using measurements taken from a short time interval. Since $g=g(A,\omega), s=s(A,\omega)$ (see equation \ref{eq:app1.3}), and the response amplitude $A$ is assumed to be a constant, it is important that the response frequency will vary between samples, to obtain different values of $g,s$ along the fitted line. Since noise affects both sides of the equation, the orthogonal line fitting method is proposed for solving the equation \cite{Shakarji1998Least-SquaresSystem.}. The formulation here accounts for zero-mean noise with similar strength in both terms.
When solving the fitting problem, good practice would be to omit parts of the measurements during which the response amplitude varies rapidly (for example, due to transient response to the sudden onset of excitation). A procedure for refining the estimation is brought in Appendix \ref{app:refineConstEst}.
The stiffness term estimation method enables solving equation \ref{eq2.6} and the results can be transformed to natural frequency and damping parameter using \ref{eq2.3}. By estimating the modal parameters and the instantaneous response amplitude of a system at every instant, the relationships $\omega_n(A),h(A)$, can be graphed without any apriori knowledge of the system, using the method we will call 'FORCEVIBmod'.
\subsection{Choice of excitation signal}\label{subsec:ExcitSig}
The choice of an appropriate excitation signal is crucial to the method's accuracy. There are a few considerations that should be taken when choosing an excitation signal. First, the signal should vary slowly: as part of the derivation of equation \ref{eq2.4}, we assumed that the modal parameters vary much slower than the excitation and response signals, i.e. – their spectral contents do not overlap (see the Bedrosian theorem \cite{Bedrosian1963ATransforms}). Since the modal parameters are functions of the instantaneous response amplitude, we need to ensure that the response amplitude varies slowly by choosing an appropriate excitation signal, that is slowly varying as well. Next, to investigate the system's behavior at a certain amplitude range, the excitation signal should be chosen such that the system will respond in the amplitude range of interest. Finally, to estimate the constant stiffness term, it is imperative that the instantaneous frequency of the response signal will vary in time. 
A compelling option for an excitation signal that fulfills all of the requirements, i.e. slow amplitude variation, response in a broad amplitude range, and frequency variation, would be a swept sine with constant amplitude and slow sweep rate.
For linear modal testing using swept sine excitation, it was shown that the sweep rate should be lower than $h^2/4$ \cite{Torvik2011OnBandwidths,Bourquard2019CommentResonance}. However, the simulations and experiments brought in the next sections show that accurate results can be obtained using much higher sweep rates.
\subsection{Method summary and comparison with the FORCEVIB method}
\begin{figure}[t]
    \centering
    \includegraphics[width=0.46\textwidth]{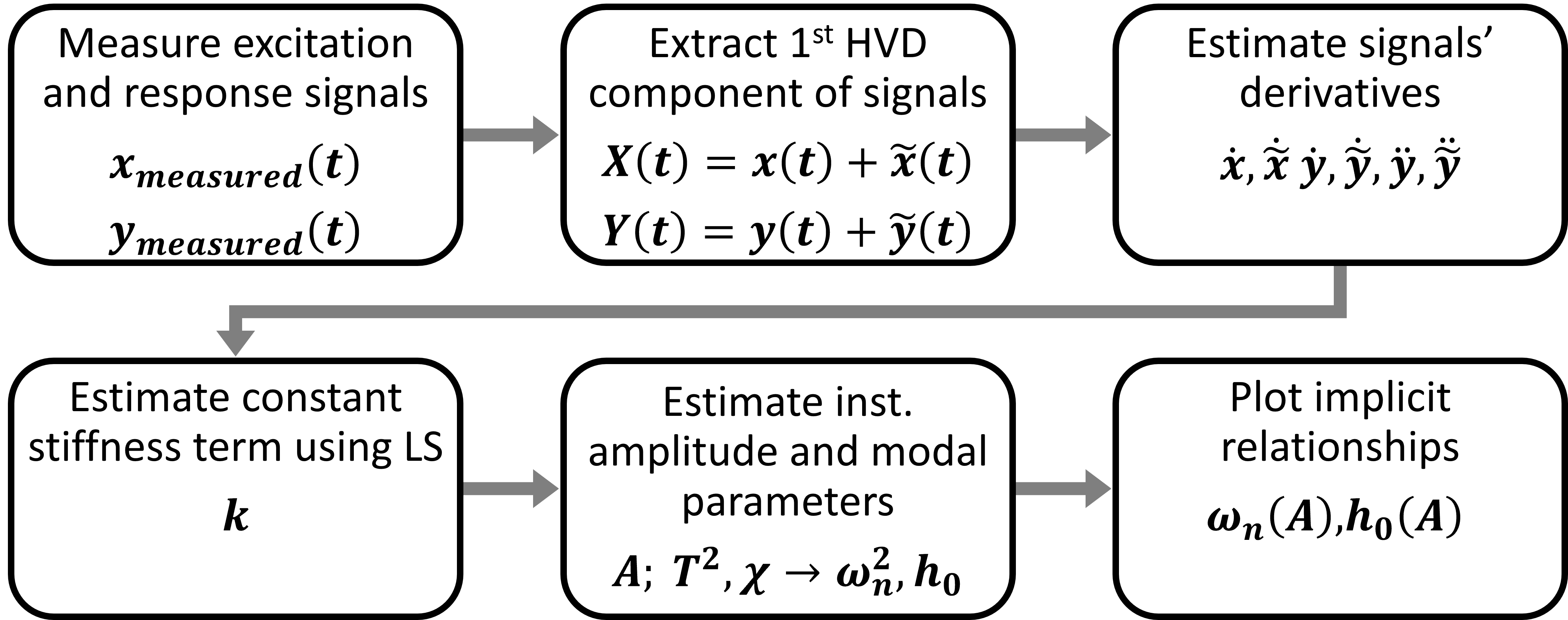}
    \caption{graphic summary of the FORCEVIBmod method}
    \label{fig:MethodSummary}
\end{figure}

To summarize this section, a step-by-step diagram of the identification process is brought in figure \ref{fig:MethodSummary}.

Additionally, a comparison between the FORCEVIB method, intended for variable-stiffness systems, and the FORCEVIBmod method, intended for variable-inertia systems is brought in table \ref{Ta:1MethodComparison}. The comparison shows the differential equations describing the systems identified by the method, the identified modal parameters, and the relations between the modal parameters and estimated signals. A discussion regarding the identification of unforced variable-mass oscillators is brought in appendix \ref{App:UnforcedIdent}
\begin{table*}[t]
\caption{Comparison between the FORCEVIB and FORCEVIBmod methods.}
\label{Ta:1MethodComparison}
    \begin{center}
        \begin{tabularx}{\textwidth}{ |l|>{\centering\arraybackslash}X|>{\centering\arraybackslash}X| }
            \hline
            \textbf{ } & \textbf{FORCEVIB} & \textbf{FORCEVIBmod}\\
            \hline
            \textbf{Analytic signal equation} & $\Ddot{Y}+2h\Dot{Y}+\omega_n^2Y=\frac{X}{m}$ & $T^2\Ddot{Y}+2\chi\Dot{Y}+Y=\frac{X}{k}$ \\
            \hline
            \textbf{Modal parameters} & $\omega_n^2=\frac{k}{m}$, Natural frequency \newline $h=\frac{c}{2m}$, Damping parameter \newline $m$, Constant parameter (mass) & $T^2=\frac{m}{k}$, Inverse natural frequency \newline $\chi=\frac{c}{2k}$, Modified damping parameter \newline $k$, Consant parameter (stiffness) \\
            \hline
            \textbf{Equations for modal parameters} & 
            $\omega _n^2 = \frac{{x\Dot\Tilde{y} - \Tilde{x}\Dot{y}}}{{m\left( {y\Dot{\Tilde{y}} - \Tilde{y}\Dot{y}} \right)}} - \frac{{\Ddot{\Tilde{y}}\Dot{y} + \Ddot{y}\Dot{\Tilde{y}}}}{{y\Dot{\Tilde{y}} - \Tilde{y}\Dot{y}}}$ &
            ${T^2} = \frac{{x\Dot{\Tilde{y}} - \Tilde{x}\Dot{y}}}{{k\left( {\Ddot{y}\Dot{\Tilde{y}} - \Dot{y}\Ddot{\Tilde{y}}} \right)}} + \frac{{\Tilde{y}\Dot{y} - y\Dot{\Tilde{y}}}}{{\Ddot{y}\Dot{\Tilde{y}} - \Dot{y}\Ddot{\Tilde{y}}}}$ \\
            \textbf{ } & 
            $2h = \frac{{\Tilde{x}y - x\Tilde{y}}}{{m\left( {y\Dot{\Tilde{y}} - \Tilde{y}\Dot{y}} \right)}} - \frac{{\Tilde{y}\Ddot{y} - \Ddot{\Tilde{y}}y}}{{y\Dot{\Tilde{y}} - \Tilde{y}\Dot{y}}}$  &
            $\chi  = \frac{{\Tilde{x}\Ddot{y} - x{\Ddot{\Tilde{y}}}}}{{2k\left( {\Ddot{y}\Dot{\Tilde{y}} - \Dot{y}\Ddot{\Tilde{y}}} \right)}} + \frac{{y\Ddot{\Tilde{y}} - \Tilde{y}\Ddot{y}}}{{2\left( {\Ddot{y}\Dot{\Tilde{y}} - \Dot{y}\Ddot{\Tilde{y}}} \right)}}$ \\ 
            \hline
        \end{tabularx}
    \end{center}
\end{table*}
\section{Method validation}\label{sec:MethodValidation}
We will now validate the proposed method by investigating a simple system with variable inertia, whose response is calculated using numerical integration. The FORCEVIBmod method will be used to identify the relationships between the system’s modal parameters and its response amplitude, and the identification results will be compared to an asymptotic approximation of these relationships.
\subsection{Simple mathematical system}
Consider a forced, undamped oscillator, whose stiffness is linear with constant $k$ and its inertia is nonlinear and symmetric with the position $y$ as:
\begin{equation}\label{eq:nonlinear oscillator}
    (m+\beta y^2)\Ddot{y}+ky=x
\end{equation}
Where $m$ is a linear inertia term, $\beta$ is a nonlinear inertia term and $x$ is excitation force.

This system does not describe any physical system, it is a mathematical construct that can be easily analyzed asymptotically. We can find the relationship between the system's instantaneous frequency and its response amplitude using the Lindstedt-Poincare' method \cite{Nayfeh1995NonlinearOscillations} and compare it with the identification results of the proposed method. The relationship between the natural frequency and the response amplitude obtained from the asymptotic approximation is:
\begin{equation}
    \begin{split}
        {\omega _n}\left( A \right) &= {\omega _0} - \frac{3}{8}{A^2}{\omega _0}\left( {\frac{\beta }{m}} \right) +\\ &\frac{{65}}{{356}}{A^4}{\omega _0}{\left( {\frac{\beta }{m}} \right)^2} + O\left( {{\omega _0}{{\left( {\frac{\beta }{m}} \right)}^3}} \right)
    \end{split}
\end{equation}
The full derivation is brought in appendix \ref{App:AsymptoticAna}. 
\subsection{Comparison to identification results}
We simulate the system's response to a chirp excitation of the form:
\begin{equation}
    x\left( t \right) = A\sin \left( {{\omega _1}t + \left( {{\omega _2} - {\omega _1}} \right)\frac{{{t^2}}}{{2T}}} \right)
\end{equation}
The numerical parameters used for the investigation are brought in table \ref{Ta:2NumPara_SimpSys}.
\begin{table}[b]
\caption{Numerical values used in the numerical investigation of the simple system.}
\label{Ta:2NumPara_SimpSys}
    \begin{center}
        \begin{tabular}{ |c|c|c| }
            \hline
            \textbf{Parameter} & \textbf{Value} & \textbf{Units}\\
            \hline
            $\omega_n=\sqrt{\frac{m}{k}}$ & $2\pi\cdot30$ & $\frac{rad}{\sec}$\\
            $\varepsilon=\frac{\beta}{m}$ & $0.1$ & $\frac{1}{m^2}$\\
            $A$ & $10^3$ & N\\
            $\omega_1$ & $2\pi\cdot20$ & $\frac{rad}{\sec}$\\
            $\omega_2$ & $2\pi\cdot40$ & $\frac{rad}{\sec}$\\
            $T$ & $10$ & sec\\ 
            \hline
        \end{tabular}
    \end{center}
\end{table}

\begin{figure*}[t]
     \centering
     \begin{subfigure}[b]{0.45\linewidth}
         \centering
         \includegraphics[width=\linewidth]{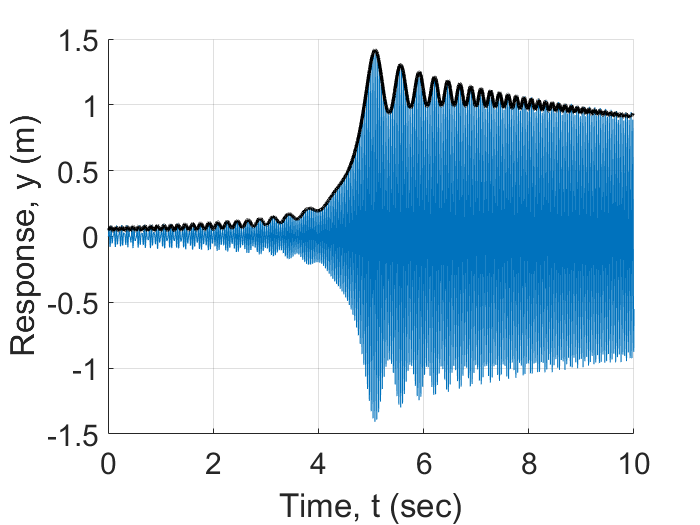}
         \caption{}
         \label{fig:SimpSys_TimeResponse}
     \end{subfigure}
     \hfill
     \begin{subfigure}[b]{0.45\linewidth}
         \centering
         \includegraphics[width=\linewidth]{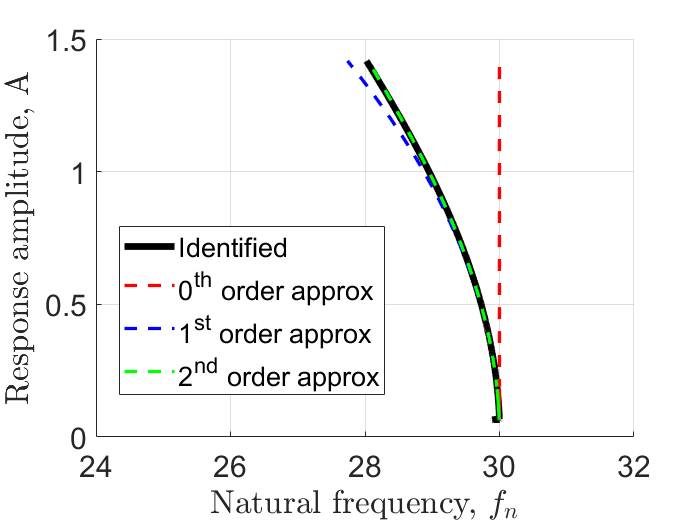}
         \caption{}
         \label{fig:SimpSys_FreqCurvewAsymApprox}
     \end{subfigure}
     \hfill
        \caption{(a) Time-history and envelope of the mathematical system's response to chirp excitation (b) Comparison between identification results (solid black line) and asymptotic approximations (dashed lines) of various orders describing the relationship between the system's natural frequency and response amplitude.}
        \label{fig:SimpSysResults}
\end{figure*}

The FORCEVIBmod method was used to identify the relationship between the instantaneous natural frequency and the response amplitude of the system from the time-history results of the numerical integration, shown in figure \ref{fig:SimpSys_TimeResponse}. First, the stiffness $k$ of the system was estimated by the means described in section \ref{subsec:ConstTermEst}, with an approximation error of 0.06\%. Next, the natural frequency and response amplitude were estimated at each instant using the relations in equation \ref{eq2.6}. The identified results and the first three orders of the asymptotic approximation are brought in figure \ref{fig:SimpSys_FreqCurvewAsymApprox}. We see that as the order of the asymptotic approximation increases, the approximations approach the identification results. This confirms that the proposed method correctly identifies the relationship between the system's modal parameters and its response amplitude. In the next section the method's identification capabilities are demonstrated for more complicated physical systems. 
\section{Case studies}\label{Case studies}
We now consider more complicated nonlinear systems with variable inertia. Since obtaining the relationships between the systems’ modal parameters and their response amplitudes analytically is not a trivial matter, identification of these relationships from simulations or experiments is a useful tool for characterizing the dynamics of these systems. 
\subsection{Electrical circuit with saturated inductor}\label{subsec:RLC}
We consider a series RLC circuit with a constant resistor  $R$, a constant capacitor $C$, and an inductor operating near saturation, i.e., its inductance $L$ depends on the current $i$ as $L(i)$. The circuit is depicted in figure \ref{fig:RLC_sim_circuit}. The differential equation describing the circuit can be derived using Kirchhoff’s voltage law:
\begin{equation}\label{eq4.1RLCeq_i}
    L(i)\frac{di}{dt}+Ri+\frac{1}{C}q=e(t)
\end{equation}
Where $q(t)$ is the capacitor’s charge and $e(t)$ is the voltage of a variable source. Since the current is the rate of change of charge, i.e. $i=\frac{dq}{dt}$ we can rewrite equation \ref{eq4.1RLCeq_i}:
\begin{equation}\label{eq4.2RLCeq_q}
    L(\Dot{q})\Ddot{q}+R\Dot{q}+\frac{1}{C}q=e(t)
\end{equation}
Equation \ref{eq4.2RLCeq_q} has the same form as equation \ref{eq2.1}, making it a viable candidate for demonstrating the FORCEVIBmod method, as the variable-inductance takes the role of variable-inertia.
Identification of an inductor's inductance-current relationship has been discussed in various works before \cite{Oliveri2022NonlinearSurvey,DiCapua2016AApplications,DiCapua2017FerriteDesign}. The method proposed in this paper may improve upon current approaches, as it is a non-parametric approach that requires simple measurements.
\subsubsection{Ideal circuit simulation}\label{subsubsec:RLC_simulation}
For demonstration purposes, we model the inductor's nonlinearity using the current-dependent inductance model proposed in \cite{DiCapua2017FerriteDesign}:
\begin{equation}
    \begin{split}
        L &= {L_{ds}} + \\ &\frac{{{L_{nom}} - {L_{ds}}}}{2}\left( {1 - \frac{2}{\pi }   {\arctan}\left( {\sigma  \cdot \left( {i - {i^*}} \right)} \right)} \right)
    \end{split}
\end{equation}

\begin{figure*}[t]
     \centering
     \begin{subfigure}[b]{0.45\linewidth}
         \centering
         \includegraphics[width=\linewidth]{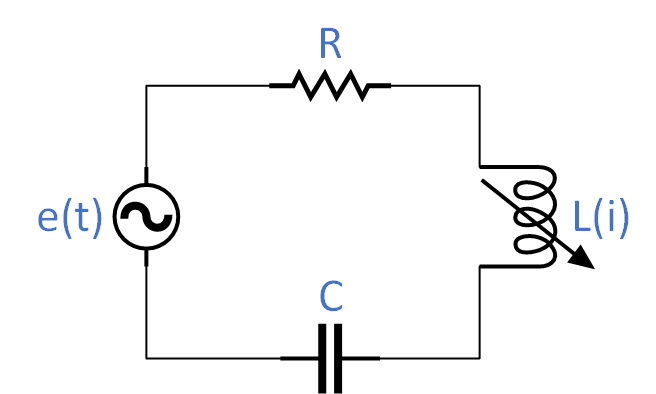}
         \caption{}
         \label{fig:RLC_sim_circuit}
     \end{subfigure}
     \hfill
     \begin{subfigure}[b]{0.45\linewidth}
         \centering
         \includegraphics[width=\linewidth]{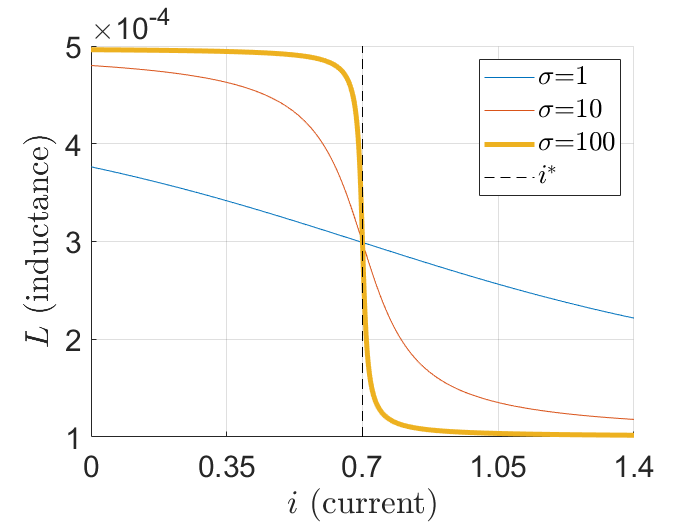}
         \caption{}
         \label{fig:RLC_sim_LvsI}
     \end{subfigure}
     \hfill
     \begin{subfigure}[b]{0.45\linewidth}
         \centering
         \includegraphics[width=\linewidth]{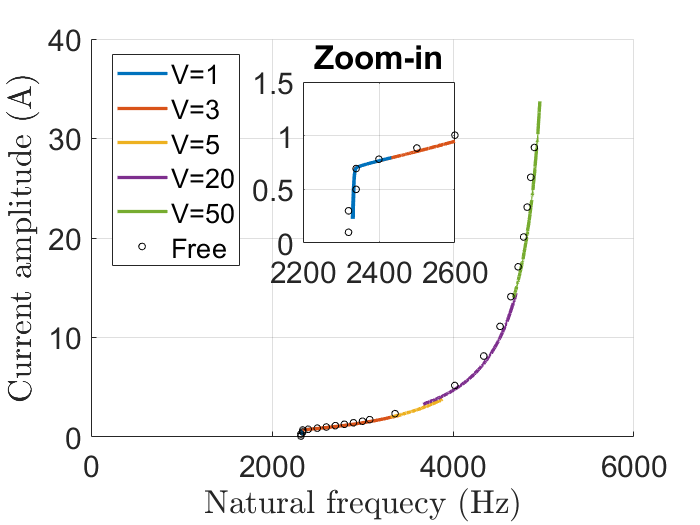}
         \caption{}
         \label{fig:RLC_sim_FreqCurve}
     \end{subfigure}
     \hfill
     \begin{subfigure}[b]{0.45\linewidth}
         \centering
         \includegraphics[width=\linewidth]{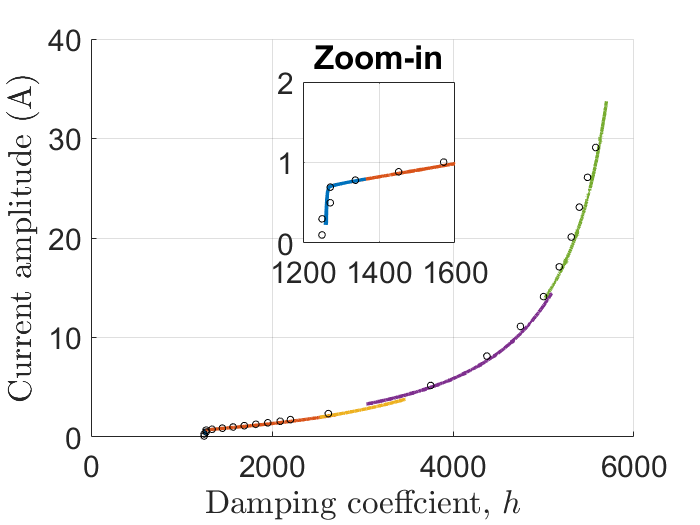}
         \caption{}
         \label{fig:RLC_sim_DampCurve}
     \end{subfigure}
     \hfill
        \caption{(a) RLC circuit with a nonlinear inductor (b) Inductor saturation model. The thicker line corresponds to the value of $\sigma$ used in numerical simulations (c-d) Relationships between (c) natural frequency, (d) damping coefficient and current amplitude obtained using the FORCEVIBmod method from the nonlinear RLC circuit simulation. Each line color corresponds to a different excitation amplitude. Black markers correspond to estimations obtained from the unforced circuit response.}
        \label{fig:RLC_sim}
\end{figure*}

Where $L_{ds},L_{nom}$ are the deep saturation inductance value and the nominal inductance values, and $\sigma,i^*$ are parameters used to define the inductance roll-off for increasing current. To demonstrate the effects of $\sigma,i^*$ on the inductance characteristic, we bring the graphs of $L(i)$ for varying values of $\sigma$ in figure \ref{fig:RLC_sim_LvsI}. We can see that $\sigma$ controls how steep the inductance transition is, while $i^*$ is the current for which the inductance does not vary with $\sigma$.

Multiplying Equation \ref{eq4.2RLCeq_q} by $C$ we obtain an equation in the form of equation \ref{eq2.2}:
\begin{equation}\label{eq4.4}
    CL\Ddot{q}+RC\Dot{q}+q=Ce(t)
\end{equation}
Comparing equation \ref{eq4.4} with equation \ref{eq2.4}, the circuit's modal parameters are $T^2=CL,2\chi=RC,k=\frac{1}{C}$ . 

Using numerical integration, we simulate the circuit's response to a swept sine excitation voltage of the form:
\begin{equation}
    e\left( t \right) = V\sin \left( {2\pi {f_1}t + 2\pi \frac{{{f_2} - {f_1}}}{{2T}}{t^2}} \right)
\end{equation}
As mentioned in section \ref{subsec:ExcitSig}, to study the circuit's response in a broad range of current amplitudes, the excitation signal should be chosen such that the circuit would respond in the entire amplitude range of interest. When exciting a system with a swept sine of constant amplitude, the variation in the system's response amplitude would not necessarily cover the entire range of response amplitudes of interest. To study the system's dynamics across the entire amplitude range, we simulate the circuit's response to a few voltage signals with various amplitudes, each covering a different portion of the desired range of response amplitudes. The different graphs can then be "stitched" together to form a graph covering the entire amplitude range of interest. The numerical parameters chosen for the electric circuit are brought in table \ref{Ta:3NumPara_RLC}.
\begin{table}
\caption{Numerical values used in the numerical investigation of the nonlinear circuit.}
\label{Ta:3NumPara_RLC}
    \begin{center}
        \begin{tabular}{ |c|c|c| }
            \hline
            \textbf{Parameter} & \textbf{Value} & \textbf{Units}\\
            \hline
            $L_{nom}$ & $498$ & $\mu$H\\
            $L_{ds}$ & $100$ & $\mu$H\\
            $i^*$ & $0.7$ & A\\
            $\sigma$ & $100$ & A$^{-1}$\\
            $C$ & $9\cdot10^{-6}$ & F \\
            $R$ & $1.25$ & $\Omega$\\ 
            $f_1$ & $1500$ & Hz\\
            $f_2$ & $4500$ & Hz\\
            $T$ &   10  &   $\sec$\\
            $V$ &   $1,3,5,20,50$ & V\\
            \hline
        \end{tabular}
    \end{center}
\end{table}

Using the FORCEVIBmod method we analyze the time-series results of the simulations to obtain estimates of the relations between the natural frequency and the current amplitude, and between the damping coefficient and current amplitude. The relations are brought graphically in figures \ref{fig:RLC_sim_FreqCurve} and \ref{fig:RLC_sim_DampCurve}, with each line color corresponding to a single simulation with a different excitation voltage amplitude. In each simulation, the system responds in a different range of amplitudes, and together all the graphs describe the variation in the modal parameters across the entire amplitude range.
Along with the identification results of the forced oscillator, estimates of the natural frequency of the undamped circuit are brought for comparison, shown in figures \ref{fig:RLC_sim_FreqCurve} and \ref{fig:RLC_sim_DampCurve} with black circular markers. Since an unforced circuit oscillates at its natural frequency, we can estimate the circuit’s natural frequency from simulations of its undamped response to initial conditions. By varying the initial conditions, we obtain oscillations with different amplitudes. Estimating the oscillation frequency for each oscillation amplitude, we obtain the relation $\omega_n(A)$. Estimations of the damping coefficient are brought as well, calculated simply as $h=\omega_n\frac{RC}{2}$. We can see that the identification results agree with the natural frequency estimations obtained from the unforced, undamped, response of the circuit. 
As expected, the natural frequency increases as the response amplitude increases, due to the decrease in the inductance. We can see that for low response amplitudes the estimated natural frequency is near $f_{nom}\frac{1}{2\pi\sqrt{L_{nom}C}}$ and for very high amplitudes it approaches the limiting value of $f_{ds}\frac{1}{2\pi\sqrt{L_{ds}C}}$. According to figure \ref{fig:RLC_sim_LvsI}, the inductance approaches its limiting value near 1A; However, it appears that the natural frequency approaches $f_{ds}$ at a much higher response amplitude. Since the inductance varies continuously during oscillations, its average value is lower than the inductance value which corresponds to the current amplitude, and since the instantaneous natural frequency is an averaged property, we can expect the natural frequency to approach a limiting value for a higher response amplitude. We can also see that the damping coefficient increases with the increase in the current amplitude, which is expected as the damping coefficient is inversely proportional to the inductance. Obtaining these relations analytically for the nonlinear circuit is rather complicated, the free vibration approach is very inefficient, and, in many practical cases, not feasible, hence non-parametric identification from forced response measurements proves useful. 
\subsubsection{Experimental demonstration}\label{subsubsec:RLCexp}
\begin{figure*}[t]
     \centering
     \begin{subfigure}[b]{0.3\linewidth}
         \centering
         \includegraphics[width=\linewidth]{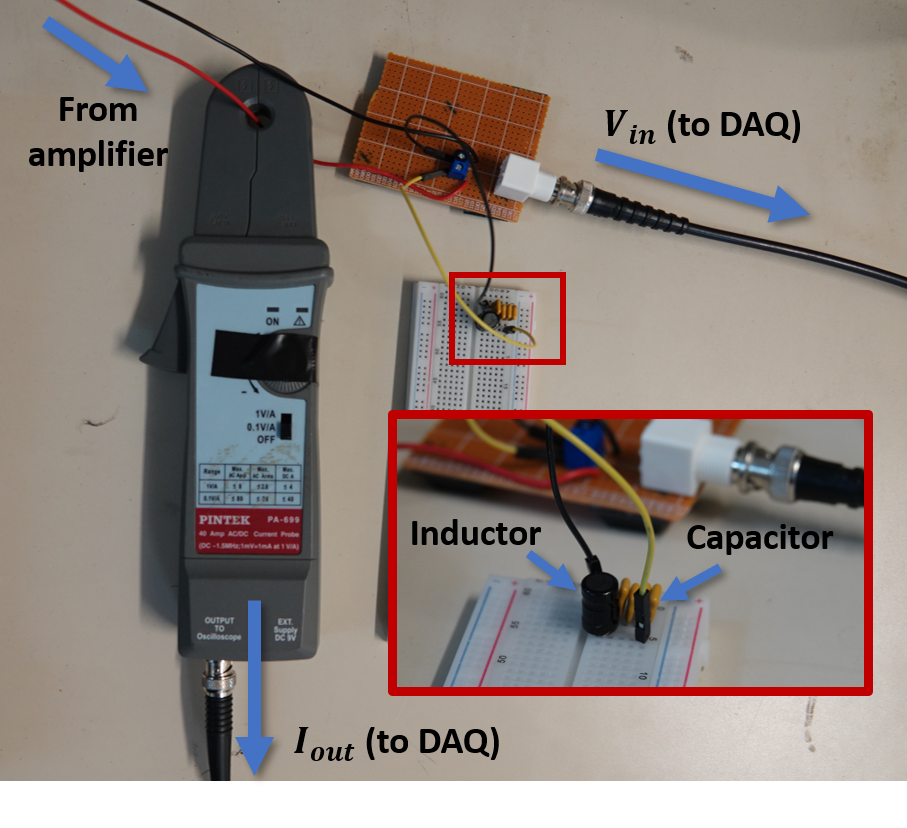}
         \caption{}
         \label{fig:RLC_exp_pic}
     \end{subfigure}
     \quad
     \begin{subfigure}[b]{0.4\linewidth}
         \includegraphics[width=\linewidth]{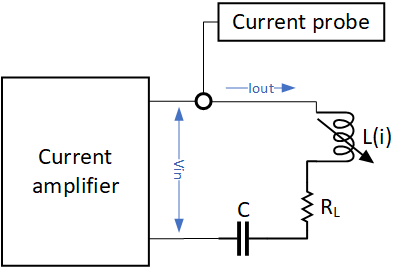}
         \caption{}
         \label{fig:RLC_exp_schematics}
     \end{subfigure}
     \hfill
     \begin{subfigure}[b]{0.7\linewidth}
         \centering
         \includegraphics[width=\linewidth]{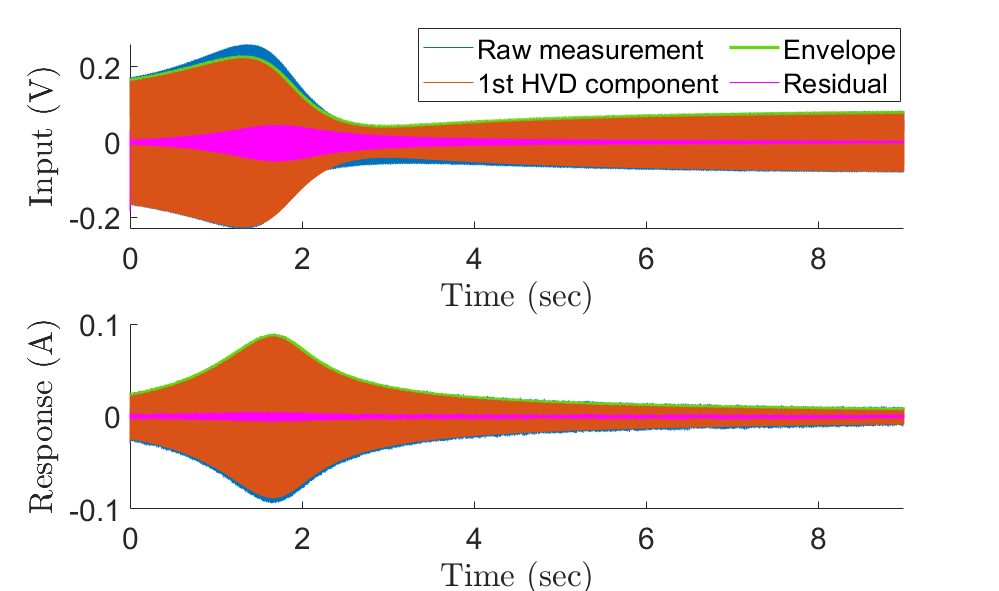}
         \caption{}
         \label{fig:RLC_exp_timeseries}
     \end{subfigure}
     \hfill
     \begin{subfigure}[b]{0.45\linewidth}
         \centering
         \includegraphics[width=\linewidth]{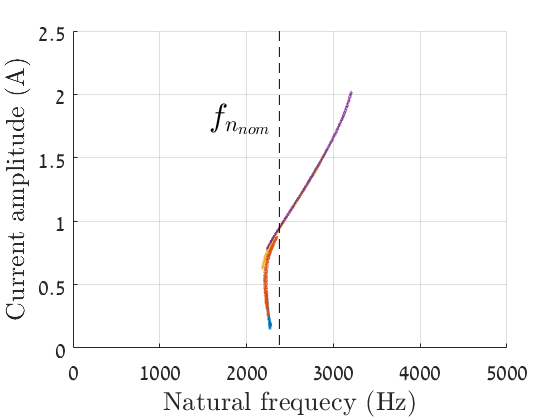}
         \caption{}
         \label{fig:RLC_exp_freqcurve}
     \end{subfigure}
     \hfill
     \begin{subfigure}[b]{0.45\linewidth}
         \centering
         \includegraphics[width=\linewidth]{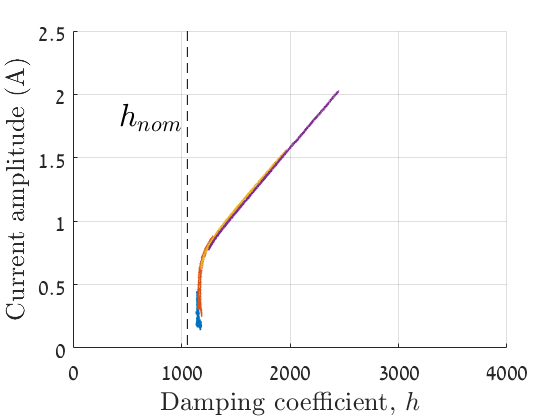}
         \caption{}
         \label{fig:RLC_exp_dampcurve}
     \end{subfigure}
     \hfill
     \hfill
        \caption{(a) Picture of RLC circuit experiment (b) Schematic description of the experiment (c) Input (top) and response (bottom) time series-measurements of RLC circuit excited with swept-sine signal with lowest amplitude. (d-e) Relationships between (d) natural frequency, (e) damping coefficient and current amplitude obtained using the FORCEVIBmod method from the RLC circuit experiment. Each line color corresponds to a different reference signal voltage amplitude. Dashed black lines correspond to estimations of nominal values at low amplitudes.}
        \label{fig:RLCExpAll}
\end{figure*}

To test the FORCEVIBmod method experimentally and demonstrate its robustness to noise, a simple LC circuit was assembled on a breadboard. The nominal circuit’s inductance, capacitance, and resistance were measured using an RLC meter at $L_{nom}=498\mu$H$,C_{nom}=9\mu$F$,R_{nom}=1.05\Omega$ respectively. From the nominal measurements, the nominal natural frequency and damping coefficient of the circuit were estimated at $f_{nom}=2377$Hz,$h_{nom}=1054\frac{1}{\sec}$.  The circuit was provided with a voltage signal $V_{in}$ of increasing frequency, provided by a current amplifier, and the current in the circuit $I_{out}$ was measured using a current probe. A picture of the RLC circuit is shown in figure \ref{fig:RLC_exp_pic}. A schematic depiction of the experimental system is shown in figure \ref{fig:RLC_exp_schematics}, with an added resistor $R_L$ accounting for the inductor’s resistance.  A more detailed description of the system is brought in appendix \ref{App:RLCdetails}.

The reference signal provided to the current amplifier was a chirp signal starting at $f_1=1500$Hz and ending at $f_2=4500$Hz whose duration was $T_{chirp}=10$sec. The experiment was repeated several times, for different chirp amplitudes. Due to the amplifier’s own dynamic response, the input voltage to the circuit did not maintain a constant voltage amplitude, but since the measured input voltage is used as part of the estimation, these effects are accounted for. The measurements from each experiment were filtered using the HVD method and the results were used to estimate the relationships between modal parameters and response amplitude using the FORCEVIBmod method. The time series measurements of the input voltage and output current for the lowest voltage amplitude are brought in figure \ref{fig:RLC_exp_timeseries} along with their first HVD component and the remaining noise. The SNR estimated for both signals is between 40 and 50db.

The identification results are brought in figures \ref{fig:RLC_exp_freqcurve} and \ref{fig:RLC_exp_dampcurve}. The capacitance value was estimated from the time-series results at the lowest amplitude at $C_{est}=9.145\mu$F, a 4.6\% error from the measured nominal value. in figures \ref{fig:RLC_exp_freqcurve} and \ref{fig:RLC_exp_dampcurve}, we can see that for low current amplitudes, the natural frequency and damping coefficient remain roughly constant, as the circuit operates in its linear regime. Their values are around $f_{n_{lin}}=2281$Hz,$h_{lin}=1171\frac{1}{\sec}$, which translates to 4\% and 11.1\% errors respectively. When the current amplitude exceeds roughly $0.7$A, the natural frequency increases rapidly. This increase is attributed to the inductor's saturation. At current amplitudes over $2$A, the rate of increase in the natural frequency appears to decrease, as the inductor approaches deep saturation. Since the damping coefficient is dependent on the inductance, it is expected to increase with the current amplitude as well. The experimental identification results were shown to be rather accurate (less than 5\% error for frequency and capacitance, errors of and less than 15\% for damping) and provided important insight into the circuit's dynamics.
\subsection{Two mass system}\label{subsec:2DOF}
\begin{figure*}[t]
     \centering
     \begin{subfigure}[b]{0.6\linewidth}
         \centering
         \includegraphics[width=\linewidth]{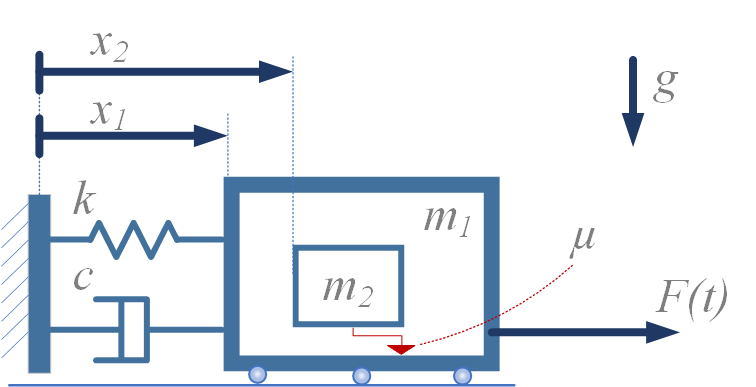}
         \caption{}
         \label{fig:2DOF_schematic}
     \end{subfigure}
     \hfill
     \begin{subfigure}[b]{0.45\linewidth}
         \centering
         \includegraphics[width=\linewidth]{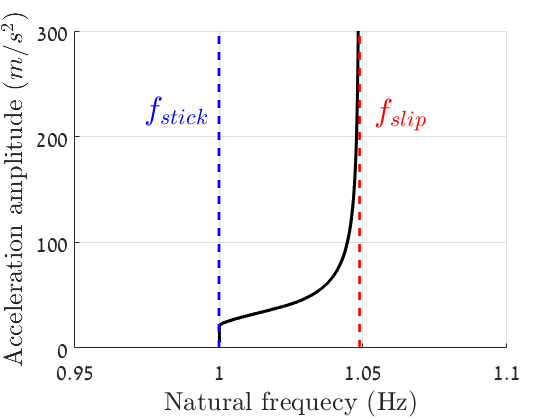}
         \caption{}
         \label{fig:2DOF_FreqCurve}
     \end{subfigure}
     \hfill
     \begin{subfigure}[b]{0.45\linewidth}
         \centering
         \includegraphics[width=\linewidth]{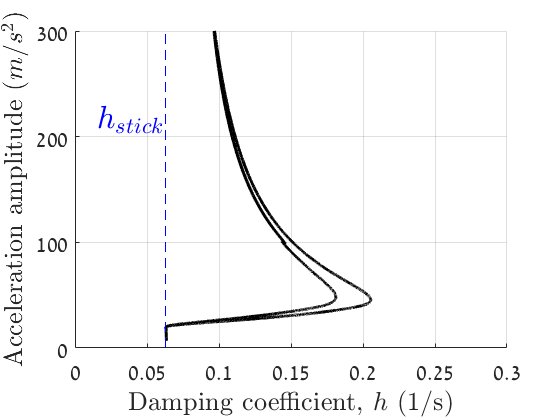}
         \caption{}
         \label{fig:2DOF_DampCurve}
     \end{subfigure}
     \hfill
        \caption{(a) Schematic description of a two-mass system with frictional interface undergoing stick-slip motion (b-c) Relationships between (b) natural frequency and vibration amplitude (c) damping coefficient and vibration amplitude obtained using the FORCEVIBmod method from the simulation of the two-mass system.}
        \label{fig:2DOF_all}
\end{figure*}

The last example of a system whose effective mass depends on its state is a two-mass system with a frictional interface, causing the masses to undergo stick-slip motion [23]. The system is depicted schematically in figure \ref{fig:2DOF_schematic}. Mass number 1 $(m_1)$ is connected to the ground with a linear spring $k$ and dashpot $c$ and its position is denoted by $x_1$. Mass number 2 $(m_2)$ is placed on the surface of $m_1$ and its position is denoted by $x_2$. The frictional interface between the masses is modeled as Coulomb friction with identical static and kinetic friction coefficients denoted as $\mu_s=\mu_k=\mu$. The system is excited by a time-varying force that acts directly on $m_1$ and the system's response is $x_1$. When the inertial forces are lower than the friction force, the masses move as a rigid body, in a regime named "stick". In the stick regime, the effective mass of the system is the sum of $m_1$ and $m_2$, hence its natural frequency and damping coefficient are $f_{stick}=\frac{1}{2\pi}\sqrt{\frac{k}{m_1+m_2}}$,$h_{stick}=\frac{c}{2(m_1+m_2)}$ respectively. When the inertial forces are larger than the friction force,  $m_2$ slides relative to $m_1$ in a regime named "slip". When the velocities of the masses are equal, the relative motion stops, and the masses return to the "stick" regime. The equations of motion of the system in each regime can be written as:
\begin{equation}
    \begin{split}
        \text{stick}: & \left\{ \begin{array}{c}
        {\left( {{m_1} + {m_2}} \right){{\Ddot{x}}_1} + c{{\Dot{x}}_1} + k{x_1} = F\left( t   \right)} \\ 
        {{{\Dot{x}}_2} = {{\Dot{x}}_1}} 
        \end{array} \right.  \\
        \text{slip}: & \left\{ \begin{array}{c}
        m_1\Ddot{x}_1 + c\Dot{x}_1 + k{x_1} = F\left( t \right) + {m_2}\mu g\operatorname{sgn} \left( {{{\Dot{x}}_2} - {\Dot{x}_1}} \right) \\ 
        m_2\Ddot{x}_2 =  - m_2\mu g\operatorname{sgn} \left( {{{\Dot{x}}_2} - {\Dot{x}}_1}    \right)
        \end{array} \right.  \\ 
    \end{split}
\end{equation}
During "slip", the friction causes dissipation and the effective mass of the system is equal to $m_1$ only. For increasing response amplitudes, the relative duration of the slip phase in a period is increased, until no stick phase is observed. Due to the increase in the relative duration of the slip phase, the effective, or, average mass of the system decreases from $m_1+m_2$ and approaches its limiting value of $m_1$.  Accordingly, the natural frequency at extremely high response amplitudes is expected to approach $f_{slip}=\frac{1}{2\pi}\sqrt{\frac{k}{m_1}}$.

The analytical treatment of such a system is rather complicated and requires piece-wise analysis, hence identifying the modal parameters of an equivalent variable-inertia system from the 2DOF system's time-series response to derive a reduced-order model may prove useful here as well. For details on the numerical simulation from which the time-series response of the system was calculated, see appendix \ref{App:2DOFdetails}. The time-series results were analyzed using the FORCEVIBmod method, and the results are brought in figures \ref{fig:2DOF_FreqCurve} and \ref{fig:2DOF_DampCurve}. The modal parameters are plotted against the system's acceleration amplitude rather than displacement amplitude, which helps in highlighting the various phenomena exhibited by the system: Up to a certain acceleration amplitude, the modal parameters do not vary with the amplitude. This behavior corresponds to the stick regime and the modal parameters are indeed equal to $f_{stick},\zeta_{stick}$. The threshold amplitude above which slip starts corresponds to the point where the inertial force exceeds the friction force. With increasing amplitude, the natural frequency increases and approaches its limiting value of $f_{slip}$. Above the threshold amplitude, the damping coefficient starts increasing as well, due to the dissipation caused by the friction forces, and keeps increasing with the response amplitude due to the increase in the relative duration of the slip phase. However, when the response amplitude increases further the damping coefficient starts decreasing. Since the magnitude of the friction force is constant, while the magnitude of the viscous damper is linear with amplitude, for high enough amplitudes the contribution of the friction force to the overall dissipation becomes negligible, hence the effective damping decreases. We can also see that the damping graph has two branches in the region of intermediate response amplitudes; Since the system is excited by a chirp force signal, different acceleration amplitudes were obtained for different frequencies. This indicates that the damping coefficient is governed mainly by the acceleration amplitudes, but other factors such as the response frequency have some effect as well, highlighting the rich dynamics of the 2DOF system.

\section{Summary and conclusions}\label{summary}
In this paper, the FORCEVIBmod method for the identification of modal parameters of variable-inertia systems was developed. The method is a modification of the FORCEVIB method which is applied for variable-stiffness systems (systems with response-dependent stiffness). It was shown that while natural frequency and damping coefficient may be used to describe the dynamics of both classes of systems, estimating the constant parameter (stiffness for variable-inertia systems and vice versa) requires slight modifications of the method.
To verify the validity of the method, a simple variable-inertia oscillator was simulated. The simulation results were analyzed using the FORCEVIBmod method and compared with an asymptotic approximation of the frequency-amplitude relationship. The analysis results were in excellent agreement with the asymptotic approximation, proving the method's validity.
To demonstrate the method's robustness to noise, the measurements of a nonlinear, oscillating RLC circuit were analyzed using the method. Despite the noise present in the measurements, the method provided a clear description of the circuit's nonlinearity caused by the inductor reaching saturation. It was shown that the FORCEVIBmod method requires only a small number of relatively quick measurements to obtain a comprehensive description of the system's dynamics, that accounts for the system's nonlinearity, without any apriori knowledge of the studied system.
Finally, a 2DOF mechanical system was discussed. Despite the system not being a classical variable-inertia system, the FORCEVIBmod method was used to derive an equivalent, reduced-order model for the system's dynamics, based on the simulation results of the system, with no analytical treatment required. 
To conclude, the FORCEVIBmod method was shown to be a useful tool for studying variable-inertia systems as it enables gaining an accurate and comprehensive description of a system's dynamics, using a relatively small number of quick measurements.
\begin{ack}     
The authors would like to thank: Dr. Michael Feldman who originally developed the FORCEVIB method and took part in reviewing the current paper, and Rakhmatulla (Roman) Shamsutdinov, who helped with the RLC experiment.
This research was supported by the Pazy Research Foundation.
\end{ack}
\appendix
\section{Assessing robustness to noise}\label{App:Noise}    
To assess the method's robustness to noise:
\begin{enumerate}
    \item The dependence of the modal parameters on the amplitude and frequency of the measured signals will be shown and the effect of noise on the identification results will be analyzed qualitatively.
    \item The effects of noise on the results will be demonstrated via simulation.
\end{enumerate}
From analytic signal theory:
\begin{equation}
    \begin{split}
        Y\left( t \right) &= A\left( t \right)\exp \left( {j\phi \left( t \right)} \right) \hfill \\
        \Dot{Y}\left( t \right) &= Y\left( t \right)\left( {\frac{{\Dot{A}\left( t \right)}}{{A\left( t \right)}} + j\omega \left( t \right)} \right) \hfill \\
        \Ddot{Y}\left( t \right) &= Y\left( t \right)\left( {\frac{{\Ddot{A}\left( t \right)}}{{A\left( t \right)}} - {\omega ^2}\left( t \right) + 2j\omega \left( t \right)\frac{{\Dot{A}\left( t \right)}}{{A\left( t \right)}} + j\Dot{\omega} \left( t \right)} \right) \hfill \\ 
    \end{split} 
\end{equation}
Substituting into equation \ref{eq2.4}:
\begin{equation}\label{eqA2.2}
    \begin{split}
        1 + &{T^2}\left( {\frac{{\Ddot{A}}}{A} - {\omega ^2}} \right) + 2\chi \frac{{\Dot{A}}}{A} \\ & + j\left( {{T^2}\left( {2\omega \frac{{\Dot{A}}}{A} + \Dot{\omega} } \right) + 2\chi \omega } \right) = \frac{1}{k}\frac{X}{Y}
    \end{split}
\end{equation}
We define $\frac{X}{Y}=\alpha+j\beta$ and separate the real and imaginary parts of \ref{eqA2.2}. Solving for $T^2,\chi$ we obtain:
\begin{equation}\label{eq:app1.3}
  \begin{split}
      T^2 = \quad & \frac{A (\beta  \Dot{A}-\alpha  A w)}{k \left(A^2 w^3+A \Dot{A} \Dot{\omega}-A \Ddot{A} w+2 \Dot{A}^2 w\right)}\\&+\frac{A^2 w}{A^2 w^3+A \Dot{A} \Dot{\omega}-A \Ddot{A} w+2 \Dot{A}^2 w} \\
      &\\
      \chi = \quad & \frac{\alpha  -A^2 \Dot{\omega}+A^2 \beta  \left(w^2\right) + 2 \alpha  A \Dot{A} w-A \beta  \Ddot{A}}{2 k \left(A^2 w^3+A \Dot{A} \Dot{\omega}-A \Ddot{A} w+2 \Dot{A}^2 w\right)}\\&-\frac{A^2 \Dot{\omega}+2 A \Dot{A} w}{2 \left(A^2 w^3+A \Dot{A} \Dot{\omega}-A \Ddot{A} w+2 \Dot{A}^2 w\right)} \\
  \end{split}
\end{equation}
Since the denominator of both expressions contains the same time-dependent expression, we should verify it is not eliminated. Since we use a slowly varying excitation signal, per the necessary assumptions used in the derivation of the proposed method, the amplitude is expected to vary slowly, i.e.:$\Ddot{A}\ll\Dot{A}\ll A$. This means that the denominator is proportional to $\sim A^2\omega^3$. Since both the response amplitude and frequency are non-zero positive values, we can expect that for slowly varying signals, the denominator will not be eliminated.
Since $A^2=y^2+\Tilde{y}^2$, the noise present in the measured signal $y$ will also affect the denominator. By using the HVD method to filter the measured signal $y$ and estimate $\Tilde{y}$ we ensure that the envelope signal is not reduced to zero due to noise.

\begin{figure}[t]
    \centering
    \includegraphics[width=0.45\textwidth]{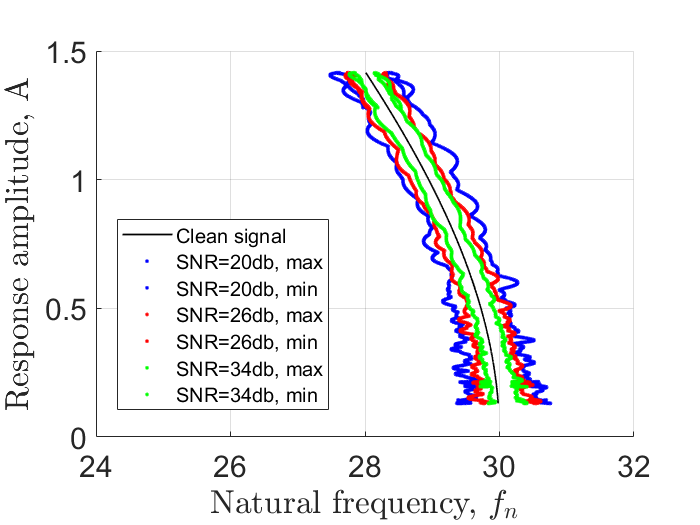}
    \caption{The effects of different SNR levels on the identification results of the system from section \ref{sec:MethodValidation}}
    \label{fig:SimSys_FreqCurveNoise}
\end{figure}

To demonstrate the method's robustness to noise, we added synthetic noise to the simulation results in section \ref{sec:MethodValidation} and analyzed the noisy measurements using the FORCEVIBmod method and the preliminary HVD step. The noise was generated from a normal distribution at three different signal-to-noise ratios (SNRs): 20db, 26db and 34db. The simulation was repeated 100 times for each noise level and the minimal and maximal frequency values obtained for each amplitude were added to the frequency plot in figure \ref{fig:SimSys_FreqCurveNoise}. We can see that the clean measurement is contained between the maximal and minimal values and that despite the relatively high noise levels, the estimation error is lower than 3\%.
\section{Refining procedure for constant term estimation}\label{app:refineConstEst}
To improve the fitting procedure described in subsection \ref{subsec:ConstTermEst}, we can use more than a single short period of time. Consider a short time window taken at some point in the measurement: since the modal parameters vary slowly in time throughout the measurement, for different windows the line intercept would be different; However, since the stiffness is constant, the slopes of all the lines are identical for all windows. This means that the stiffness term can be identified by fitting a straight line to the measurements of g,s at many time windows. The optimization problem for the stiffness term identification can be formulated as follows:

Define $N$ as the entire length of the time-series measurement. Take $m\ll N$ measurements, (i.e. a short period), and calculate $g,s$ for all $m$ measurements. Since we are not interested in fitting the intercept, remove mean value: $\Tilde{g}_j=g_j-\frac{1}{m}\sum_{i=1}^m{g_i}, \tilde{s}_j=s_j-\frac{1}{m}\sum_{i=1}^m{s_i}$. Define the distance of each point from a straight-line whose slope is k and its intercept is 0 is $d_j=e(k)\times(s_j,g_j)$, where $e(k)$ is the direction cosine specifying the orientation of the line and $\times$ denotes a vector cross product \cite{Shakarji1998Least-SquaresSystem.}. The objective function of the optimization problem would be to find the value of k that minimizes:
\begin{equation}
\mathop {\min }\limits_k \quad J = \sum\limits_i^N {{{\left| {{{\sum\nolimits_{j = i}^{i + m} {{d_j}\left( k \right)} }^2}} \right|}^2}}
\end{equation}
\section{Free variable-inertia system identification}\label{App:UnforcedIdent}
The modal parameters used to characterize the variable-inertia system are its natural frequency and damping coefficient, both of which are used to characterize variable-stiffness systems as well. Because of that, one may consider using the FORCEVIB method to characterize variable inertia systems. However,  the right-hand side of the equation describing forced variable-inertia contains the term $\frac{1}{k}$, while the right-hand side of the equation for variable-stiffness systems contains the term $\frac{1}{m}$., see Equations \ref{eq1.5} and \ref{eq2.2}. The FORCEVIBmod method is required for variable-inertia system identification because the constant parameter estimation method is slightly different for variable-stiffness and variable-inertia systems. 
For free vibration, the right-hand side is nullified and there is no need to estimate the constant parameter. An identification method for freely vibrating systems using similar ideas was previously proposed by Feldman - FREEVIB \cite{Feldman1994Non-linearFreevib}. Assuming the nonlinear terms are not zero, for free vibrations, with proper normalization, both variable-inertia and variable-stiffness systems can be described by the equation:
\begin{equation}
    \Ddot{y}(t)+h(y,\Dot{y})\Dot{y}(t)+\omega_n^2(y,\Dot{y})=0
\end{equation}
The FREEVIB method allows for the identification of $h,\omega_n$ and can be applied to both variable-inertia and variable-stiffness systems. The FREEVIB method is very useful for systems with light damping whose free response can be measured over a significant duration, but in many cases, the free vibrations cannot be measured. In these situations, the FORCEVIB and FORCEVIBmod methods are required, depending on the type of the studied system (variable-stiffness or variable-inertia).
\section{Asymptotic analysis of inertia-variable oscillator}\label{App:AsymptoticAna} 
Consider the system in section \ref{subsec:Variable-inertia systems}, described by equation \ref{eq:nonlinear oscillator}. We wish to approximate its natural frequency as a function of its response amplitude using the Lindstedt-Poincare' method. We start by normalizing the free system to the form:
\begin{equation}
    \left( 1+\varepsilon y^2 \right)\Ddot{y}+\omega_0^2 y = 0
\end{equation}
With $\varepsilon=\beta/m$ and $\omega_0^2=k/m$. Next, introduce a new time scale $\tau=\omega t$ where $\omega$ is the natural frequency of the system and $\omega_0$ is the leading order approximation of the system's natural frequency for $\varepsilon\rightarrow0$. Changing variables using the chain rule, we obtain:
\begin{equation}\label{eq:app5.3}
    \omega^2\left(1+\varepsilon y^2\right)y''+\omega_0^2y=0
\end{equation}
Here we denote $()'=\frac{d}{d\tau}()$. Expanding the response and the natural frequency to a power series in $\varepsilon$:
\begin{equation}
    \begin{split}
        y&=y_0+\varepsilon y_1 + \varepsilon^2 y_2 +...\\
        \omega &= \omega_0 + \varepsilon\omega_1 +\varepsilon^2 \omega_2 + ...\\
    \end{split}
\end{equation}
Substituting into equation \ref{eq:app5.3} and equating coefficients of $\varepsilon$ to zero leads to the following set of equations:
\begin{equation}
    \begin{split}
        \varepsilon^0: & \quad y_0''+y_0=0\\
        \varepsilon^1: & \quad y_1''+y_1=-2\left(\frac{\omega_1}{\omega_0}+y_0^2\right)y_0''\\
        \varepsilon^2: & \begin{split}
            \quad y_2''+y_2=&-2\left(\frac{\omega_1^2}{\omega_0^2}+\frac{2\omega_2}{\omega_0}+\frac{2\omega_1}{\omega_0}y_0^2+2y_0y_1\right)y_0''\\&-\left(\frac{2\omega_1}{\omega_0}+y_0^2\right)y_1''
        \end{split}\\
        \vdots         &  \\
    \end{split}
\end{equation}
The solution for the 0\textsuperscript{th}-order expansion is of the form:
\begin{equation}
    y_0=A\cos(\tau_\phi)
\end{equation}

Substituting into the 1\textsuperscript{st} order expansion, we obtain:
\begin{equation}
    \begin{split}
        {y''_1} + {y_1} &= \frac{1}{4}\left({3{A^3}\omega_0^2 + 8A{\omega _1}{\omega_0}} \right)\cos \left({\tau  + \phi} \right) \\
        & + \frac{1}{4}{A^3}{\omega_0}^2\cos \left({3\tau  + 3\phi} \right)
    \end{split}
\end{equation}
To eliminate the secular term $\cos \left({\tau  + \phi} \right)$, we require that its coefficient equals zero:
\begin{equation}
    \left({3{A^3}\omega_0^2 + 8A{\omega _1}{\omega_0}} \right)
\end{equation}
and obtain the 1\textsuperscript{st}-order approximation for the system's instantaneous frequency:
\begin{equation}
    \omega_1=\frac{3}{8}A^2\omega_0
\end{equation}
We can now solve the equation for the 1\textsuperscript{st}-order expansion. Following the same procedure, we can obtain the 2\textsuperscript{nd}-order approximation for the instantaneous frequency to be:
\begin{equation}
    \omega_2=\frac{65}{256}^4\omega_0
\end{equation}
And we can continue to the desired order of approximation. 
\section{Detailed description of the RLC circuit experiment}\label{App:RLCdetails} 
\begin{figure*}[t]
    \centering
    \includegraphics[width=\linewidth]{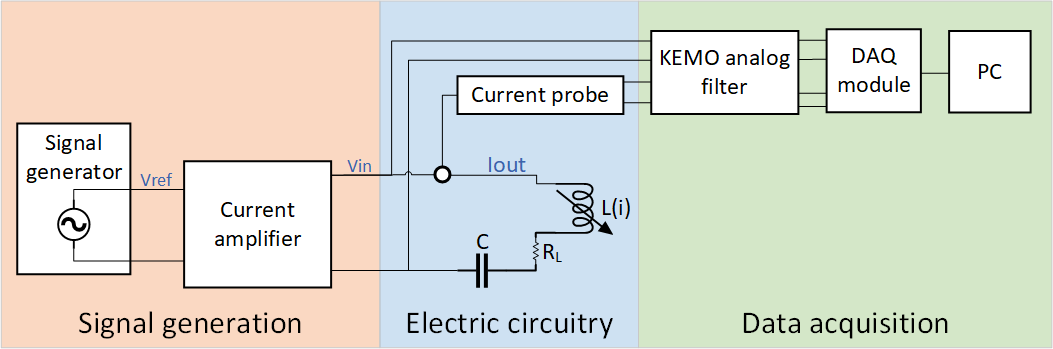}
    \caption{Detailed schematic description of the RLC experimental system}
    \label{fig:RLCFull}
\end{figure*}
The FORCEVIBmod method was demonstrated experimentally by investigating an RLC circuit, as described in section  \ref{subsubsec:RLCexp}. The RLC circuit was excited with a voltage signal $V_{in}$ of increasing frequency using a Labworks PA-151 current amplifier. The reference signal $V_{ref}$ to the amplifier was generated using an Agilent 33220A arbitrary waveform generator. The current $I_{out}$ was measured using a Pintek PA-699 current probe. The voltage provided by the current amplifier and the current signal from the current probe were filtered through a KEMO analog filter, model vbf10m, set as a low-pass filter with a $7$kHz cut-off frequency. The filtered signals were measured using a data-translation DT9873C data acquisition module connected to a PC. A schematic depiction of the experimental system is shown in Figure \ref{fig:RLCFull}.
\section{Numerical simulation of a stick-slip system}\label{App:2DOFdetails} 
The equations of motion of the two-mass system described in section \ref{subsec:2DOF} and depicted in figure \ref{fig:2DOF_schematic} are formulated using the Lagrange method with constraints. The vector of degrees of freedom:
\begin{equation}
    q = {\left( {\begin{array}{{c c}}
    {{x_1}}&{{x_2}}
    \end{array}} \right)^T}
\end{equation}
The sums of kinetic energies $T$, potential energies $V$, and the dissipation function $D$ are:
\begin{equation}
    \begin{split}
         T&=\frac{1}{2}m_1x_1^2+\frac{1}{2}m_2x_2^2 \\ V&=\frac{1}{2}k x_1^2 \\ D&=\frac{1}{2}c\Dot{x}_1^2
    \end{split}
\end{equation}
The friction force is accounted for by introducing kinematic constraint, stating the mass's velocities are identical, i.e.:
\begin{equation}
    \Dot{x}_1=\Dot{x}_2
\end{equation}
The constraint can be expressed using the holonomic matrix 
$W=\left( \begin{array}{cc}  1&-1  \\ \end{array} \right)$ as:
\begin{equation}\label{eq:app5.14}
    W\Dot{q}=0
\end{equation}
Substituting the above equations into the constrained Lagrange equation
\begin{equation}
    \begin{split}
      \frac{d}{{dt}}\left( {\frac{{dT}}{{d{{\Dot{q}}_i}}}} \right) - \frac{{dT}}{{d{q_i}}} + \frac{{dD}}{{d{{\Dot{q}}_i}}} + \frac{{dV}}{{d{q_i}}} = \\ = {Q_i} + \sum\limits_{j = 1}^{{n_{constraints}}} {{w_{ij}}\left( q \right){\lambda _j}}
    \end{split}
\end{equation}
yields the system’s constrained equations of motion:
\begin{equation}\label{eq:app5.16}
    \underbrace {\left( {\begin{array}{{c c}}
  {m_1} & 0 \\ 
  0 & {m_2} 
\end{array}} \right)}_M\Ddot{q} +
\underbrace {\left( {\begin{array}{{c c}}
  c & 0 \\ 
  0 & 0 
\end{array}} \right)}_C\Dot{q} +
\underbrace {\left( {\begin{array}{{c c}}
  k & 0 \\ 
  0 & 0 
\end{array}} \right)}_K q = 
\underbrace {\left( {\begin{array}{{c c}}
  {F\left( t \right)} \\ 
  0 
\end{array}} \right)}_Q +
\underbrace {\left( {\begin{array}{{c c}}
  1 \\ 
  { - 1} 
\end{array}} \right)}_{{W^T}}\lambda
\end{equation}
The Lagrange multiplier $\lambda$ represents the friction force that constrains the motion of the two masses together. Since the time-derivative of the constraint equation \ref{eq:app5.14} is zero, it can be used to calculate the value of the friction force as a function of the system's state:
\begin{equation}
    \frac{d}{dt}(W\Dot{q})=0\rightarrow \Dot{W}\Dot{q}+W\Ddot{q}=0
\end{equation}
substituting equation \ref{eq:app5.16} gives:
\begin{equation}
    W{M^{ - 1}}\left( {{W^T}\lambda  + Q - C\Dot{q} - Kq} \right) = 0
\end{equation}
Solving for the constraint force $\lambda$:
\begin{equation}\label{eq:app5.19}
    \lambda  = {\left( {W{M^{ - 1}}{W^T}} \right)^{ - 1}}\left( {W{M^{ - 1}}\left( {C\Dot{q} + Kq - Q} \right)} \right)
\end{equation}
which is the friction force required for the masses to keep sticking together. 
The model assumes Coulomb friction, i.e. the friction force is limited by $|f|\leq\mu N$. This means that when the calculated constraint force exceeds the friction limit, the constraint no longer applies, and the system should transition from stick to slip. When the masses' velocities are equal and the constraint force is below the friction limit, the system transitions back from slip to stick.
To account for the transitions in the numerical integration scheme, the constraint force $\lambda$ in equation \ref{eq:app5.16} is replaced by a friction force $f$. During stick, when the constraint force magnitude is lower than $\mu N$, the friction force $f$ is equal to the constraint force $\lambda$. During slip, the friction force $f$ is equal to $\mu N$ with the appropriate direction. With this change, the two-mass system is simulated by numerical integration of the equation:
\begin{equation}
M\Ddot{q} + C\Dot{q} + Kq = Q + {W^T}f
\end{equation}
While the friction force $f$ is continuously calculated using the following IF statement:
\begin{equation}
    \begin{array}{l l l}
         if & 
         \left|\lambda\right|>\mu N\vee\left|{\Dot{x}_2-\Dot{x}_1}\right| > tol &
         f=\mu N\operatorname{sgn} \left(\Dot{x}_2 - \Dot{x}_1 \right) \\
         else & & f=\lambda
    \end{array}
\end{equation}
And the value of $\lambda$ is continuously calculated using equation \ref{eq:app5.19}.
The IF statement used for the friction force $f$ calculation states that the system should be in slip mode if either the magnitude of the constraint force $\lambda$ is larger than $\mu N$ or the masses' velocities are not equal. 
The velocity condition is important because during slip, equation \ref{eq:app5.19} no longer applies, yet it is continuously calculated and might yield a value of $\lambda$ lower than $\mu N$. To prevent the system from transitioning back to stick, the condition $\left|{\Dot{x}_2-\Dot{x}_1}\right| > tol$ is introduced, with $tol$ being some numerical tolerance. The transition from slip to stick is allowed only when the mass's velocities are identical and when the magnitude of the constraint force is lower than $\mu N$.
\bibliographystyle{unsrt}
\bibliography{IdentMethod}

\begin{thebibliography}{10}

\bibitem{DynMechSysVarMass}
Hans Irschik and Alexander~K. Belyaev, editors.
\newblock {\em {Dynamics of Mechanical Systems with Variable Mass}}, volume 557 of {\em CISM International Centre for Mechanical Sciences}.
\newblock Springer Vienna, Vienna, 2014.

\bibitem{Cveticanin2022DynamicsMass}
L.~Cveticanin.
\newblock {Dynamics of Machines with Variable Mass}.
\newblock {\em Dynamics of Machines with Variable Mass}, 2 2022.

\bibitem{Pesce2014SystemsPosition}
Celso~Pupo Pesce and Leonardo Casetta.
\newblock {Systems with mass explicitly dependent on position}.
\newblock {\em CISM International Centre for Mechanical Sciences, Courses and Lectures}, 557:51--106, 2014.

\bibitem{Nayfeh1995NonlinearOscillations}
Ali~Hasan Nayfeh and Dean~T. Mook.
\newblock {Nonlinear Oscillations}.
\newblock {\em Nonlinear Oscillations}, 5 1995.

\bibitem{Oliveri2022NonlinearSurvey}
Alberto Oliveri, Matteo Lodi, and Marco Storace.
\newblock {Nonlinear models of power inductors: A survey}.
\newblock {\em International Journal of Circuit Theory and Applications}, 50(1):2--34, 1 2022.

\bibitem{Marinca2010DeterminationMethod}
Vasile Marinca and Nicolae Heri{\c{s}}anu.
\newblock {Determination of periodic solutions for the motion of a particle on a rotating parabola by means of the optimal homotopy asymptotic method}.
\newblock {\em Journal of Sound and Vibration}, 329(9):1450--1459, 4 2010.

\bibitem{Venkatesan1997NonlinearSystems}
A.~Venkatesan and M.~Lakshmanan.
\newblock {Nonlinear dynamics of damped and driven velocity-dependent systems}.
\newblock {\em Physical Review E}, 55(5):5134, 5 1997.

\bibitem{vanHorssen2006OnMass}
W.~T. van Horssen, A.~K. Abramian, and {Hartono}.
\newblock {On the free vibrations of an oscillator with a periodically time-varying mass}.
\newblock {\em Journal of Sound and Vibration}, 298(4-5):1166--1172, 12 2006.

\bibitem{vanHorssen2010OnMass}
W.~T. van Horssen, O.~V. Pischanskyy, and J.~L.A. Dubbeldam.
\newblock {On the forced vibrations of an oscillator with a periodically time-varying mass}.
\newblock {\em Journal of Sound and Vibration}, 329(6):721--732, 3 2010.

\bibitem{Kerschen2006PastDynamics}
Gaëtan Kerschen, Keith Worden, Alexander~F. Vakakis, and Jean~Claude Golinval.
\newblock {Past, present and future of nonlinear system identification in structural dynamics}.
\newblock {\em Mechanical Systems and Signal Processing}, 20(3):505--592, 4 2006.

\bibitem{Brunton2016DiscoveringSystems}
Steven~L. Brunton, Joshua~L. Proctor, J.~Nathan Kutz, and William Bialek.
\newblock {Discovering governing equations from data by sparse identification of nonlinear dynamical systems}.
\newblock {\em Proceedings of the National Academy of Sciences of the United States of America}, 113(15):3932--3937, 4 2016.

\bibitem{Lyons2011UnderstandingLyons.}
Richard~G Lyons.
\newblock {\em {Understanding digital signal processing / Richard G. Lyons.}}
\newblock Prentice Hall, Upper Saddle River, N.J, 3rd ed. edition, 2011.

\bibitem{Feldman1994Non-linearForcevib}
Michael Feldman.
\newblock {Non-linear system vibration analysis using Hilbert transform--II. Forced vibration analysis method 'Forcevib'}.
\newblock {\em Mechanical Systems and Signal Processing}, 8(3):309--318, 5 1994.

\bibitem{Bedrosian1963ATransforms}
E.~Bedrosian.
\newblock {A product theorem for Hilbert transforms}.
\newblock {\em Proceedings of the IEEE}, 51(5):868--869, 1963.

\bibitem{Feldman2011HilbertVibration}
Michael Feldman.
\newblock {\em {Hilbert Transform Applications in Mechanical Vibration}}.
\newblock Wiley, 2011.

\bibitem{Feldman2006Time-varyingTransform}
Michael Feldman.
\newblock {Time-varying vibration decomposition and analysis based on the Hilbert transform}.
\newblock {\em Journal of Sound and Vibration}, 295(3-5):518--530, 8 2006.

\bibitem{Xia2021ModalStructures}
Yingzhi Xia, Hui Li, Zhezhe Fan, and Jiyong Xiao.
\newblock {Modal Parameter Identification Based on Hilbert Vibration Decomposition in Vibration Stability of Bridge Structures}.
\newblock {\em Advances in Civil Engineering}, 2021, 2021.

\bibitem{Shakarji1998Least-SquaresSystem.}
Craig~M Shakarji.
\newblock {Least-Squares Fitting Algorithms of the NIST Algorithm Testing System.}
\newblock {\em Journal of research of the National Institute of Standards and Technology}, 103(6):633--641, 1998.

\bibitem{Torvik2011OnBandwidths}
Peter~J. Torvik.
\newblock {On estimating system damping from frequency response bandwidths}.
\newblock {\em Journal of Sound and Vibration}, 330(25):6088--6097, 12 2011.

\bibitem{Bourquard2019CommentResonance}
Claire Bourquard and Nicolas Noiray.
\newblock {Comment on “Slow passage through resonance”}.
\newblock {\em Physical Review E}, 100(4):047001, 10 2019.

\bibitem{DiCapua2016AApplications}
Giulia Di~Capua and Nicola Femia.
\newblock {A novel method to predict the real operation of ferrite inductors with moderate saturation in switching power supply applications}.
\newblock {\em IEEE Transactions on Power Electronics}, 31(3):2456--2464, 3 2016.

\bibitem{DiCapua2017FerriteDesign}
G.~Di~Capua, N.~Femia, K.~Stoyka, M.~Lodi, A.~Oliveri, and M.~Storace.
\newblock {Ferrite inductor models for switch-mode power supplies analysis and design}.
\newblock {\em SMACD 2017 - 14th International Conference on Synthesis, Modeling, Analysis and Simulation Methods and Applications to Circuit Design}, 7 2017.

\bibitem{Feldman1994Non-linearFreevib}
Michael Feldman.
\newblock {Non-linear system vibration analysis using Hilbert transform--I. Free vibration analysis method 'Freevib'}.
\newblock {\em Mechanical Systems and Signal Processing}, 8(2):119--127, 3 1994.

\end{thebibliography}
\end{document}